\documentclass[%
amsmath,amssymb,
aps,
]{revtex4-2}

\usepackage{graphicx}
\usepackage{dcolumn}
\usepackage{bm}
\usepackage{url}
\usepackage{xcolor}
\usepackage{titlesec}
\usepackage{caption}
\usepackage{multirow}
\usepackage{ulem}

\titlespacing*{\section}
{0pt}{4ex}{2ex}
\titlespacing*{\subsection}
{0pt}{4ex}{2ex}
\titlespacing*{\subsubsection}
{0pt}{4ex}{2ex}

\begin{document}

\title{Super-resolution GANs of randomly-seeded fields}
\author{Alejandro Güemes$^{\text{1}}$, Carlos Sanmiguel Vila$^{\text{1,2}}$, Stefano Discetti$^{\text{1}}$}
\affiliation{1: Aerospace Engineering Research Group, Universidad Carlos III de Madrid, Leganés, Spain \\
2: Sub-Directorate General of Terrestrial Systems, Spanish National Institute for Aerospace Technology (INTA), San Martín de la Vega, Spain
}%

\begin{abstract}
Reconstruction of field quantities from sparse measurements is a problem arising in a broad spectrum of applications. This task is particularly challenging when the mapping between  sparse measurements and field quantities is performed in an unsupervised manner.  Further complexity is added for moving sensors and/or random on-off status. Under such conditions, the most straightforward solution is to interpolate the scattered data onto a regular grid. However, the spatial resolution achieved with this approach is ultimately limited by the mean spacing between the sparse measurements. In this work, we propose a  super-resolution generative adversarial network (GAN) framework to estimate field quantities from random sparse sensors without needing any full-field high-resolution training. The algorithm exploits random sampling to provide incomplete views of the {high-resolution} underlying distributions. It is hereby referred to as RAndomly-SEEDed super-resolution GAN (RaSeedGAN). The proposed technique is tested on synthetic databases of fluid flow simulations, ocean surface temperature distributions measurements, and particle image velocimetry data of a zero-pressure-gradient turbulent boundary layer. The results show  excellent performance even in cases with high sparsity or with levels of noise. To our knowledge, this is the first  GAN algorithm for full-field  high-resolution estimation from randomly-seeded fields with no need of full-field high-resolution representations.
\end{abstract}

\maketitle

Sparse observations are often the only option for geophysical \cite{bolton2019applications}, astrophysical \cite{akiyama2019first}, biological \cite{yakhot2007reconstruction}, or fluid mechanical systems characterization \cite{manohar2018data,cortina2020sparse,fukami2021global}. Meteorological or oceanographic flows are remarkable examples in which atmospheric pressures, temperatures, and wind are only measured at a limited number of stations \cite{gundersen2021semi}; cardiac blood or flow control applications \cite{shen2015missing,callaham2019robust} suffer similar issues, being sensor intrusiveness and noise contamination particularly relevant. 

Various approaches to obtain full-field description from incomplete sparsely-seeded measurements can be followed depending on the sparsity level and the number of the available samples. For limited gappyness and a sufficiently large number of samples, reconstruction techniques based on extracting flow features using methodologies such as the Proper Orthogonal Decomposition (POD) \cite{berkooz1993proper} or the Dynamic Mode Decomposition (DMD) \cite{schmid2010dynamic} can be applied. For instance, the Gappy-POD filling procedures \cite{everson1995karhunen,venturi2004gappy,raben2012adaptive} estimate non-gappy POD modes and time coefficients by least-squares regression on the available sparse fields; the estimated modes are later used to calculate the reconstructed full fields. A recent approach, proposed in \citep{cortina2020sparse}, merges Gappy-POD and ensemble-particle-averaging techniques, paving the way to the reconstruction of fields with significantly higher gappyness fraction. The main advantage is that a high-resolution dictionary is not required beforehand and is constructed directly using low-resolution spatially-averaged fields and sparse fields. On the downside, a relatively high number of samples is required to reach suitable convergence. For cases with a limited number of samples, compressed-sensing (CS) methods provide efficient reconstruction using sparse representations \cite{huang2013compressive,manohar2018data,callaham2019robust}. CS is based on solving an under-determined linear system, which relates the sparse measurements with the entire fields, with additional constraints typically enforcing the smoothness of the solution \cite{manohar2018data}. A significant drawback is that they rely on a linear mapping between the sparse measurements and the high-resolution flow field. Therefore, in the cases where the relationship is strongly nonlinear or when a high signal-to-noise ratio corrupts the available measurements, the use of these methods is limited to cases with relatively low gappyness fraction. 

In the last years, neural-network-based approaches have emerged to overcome these limitations \cite{gundersen2021semi,MaulikPRF2020,fukami2021global,gao2021super,erichson2020shallow,sun2020physics,Arzani2021blood,brunton2020machine}. Neural networks have been proved to be a powerful non-linear mapping tool. Super-resolution architectures based on generative adversarial networks (GANs) have been successfully applied in turbulent flows \cite{fukami2019super,kim2020unsupervised,guemes2021coarse,deng2019super} and climate data \cite{stengel2020adversarial} to enhance the spatial resolution and/or fill gaps between sparse measurements. While the success of these works is undeniable, they are all limited by the need for a non-sparse target for training, i.e., a library of examples where the entire field is required to train the reconstruction algorithm. This requirement often limits their applicability in real cases, in which the impossibility to record high-resolution measurements for training is more the rule than the exception. 

In response to this problem, a novel super-resolution GAN framework to estimate field quantities without needing full-resolution fields for training is presented. The main advantage of the presented methodology is that it only requires the information of randomly located sensors at a variable location from sample to sample. This situation is quite common in particle-based measurements, such as tracers in particle tracking velocimetry (PTV) or phosphor thermography, but also in the case of moving sensors, such as buoys or ships. Low-resolution snapshots created by a simple binning procedure and high-resolution sparse snapshots are used for the training process, making it unnecessary to have complete high-resolution fields as the target for training. Since the proposed framework exploits the random sampling to provide incomplete views of the underlying high-resolution fields, the method is named RAndomly-SEEDed super-resolution GAN (RaSeedGAN). A set of challenging validation cases which include numerical simulations of a fluidic pinball flow, a turbulent channel flow and data from the NOAA sea surface temperature database and a zero-pressure-gradient (ZPG) turbulent boundary layer particle image velocimetry (PIV) experiment is employed to test the robustness of the methodology in complex cases where nonlinearities and noise are present.

\section*{\textcolor{black}{The Randomly-Seeded super-resolution GAN}}

\begin{figure}[t!]
        \centering
    \includegraphics[width=0.9\textwidth]{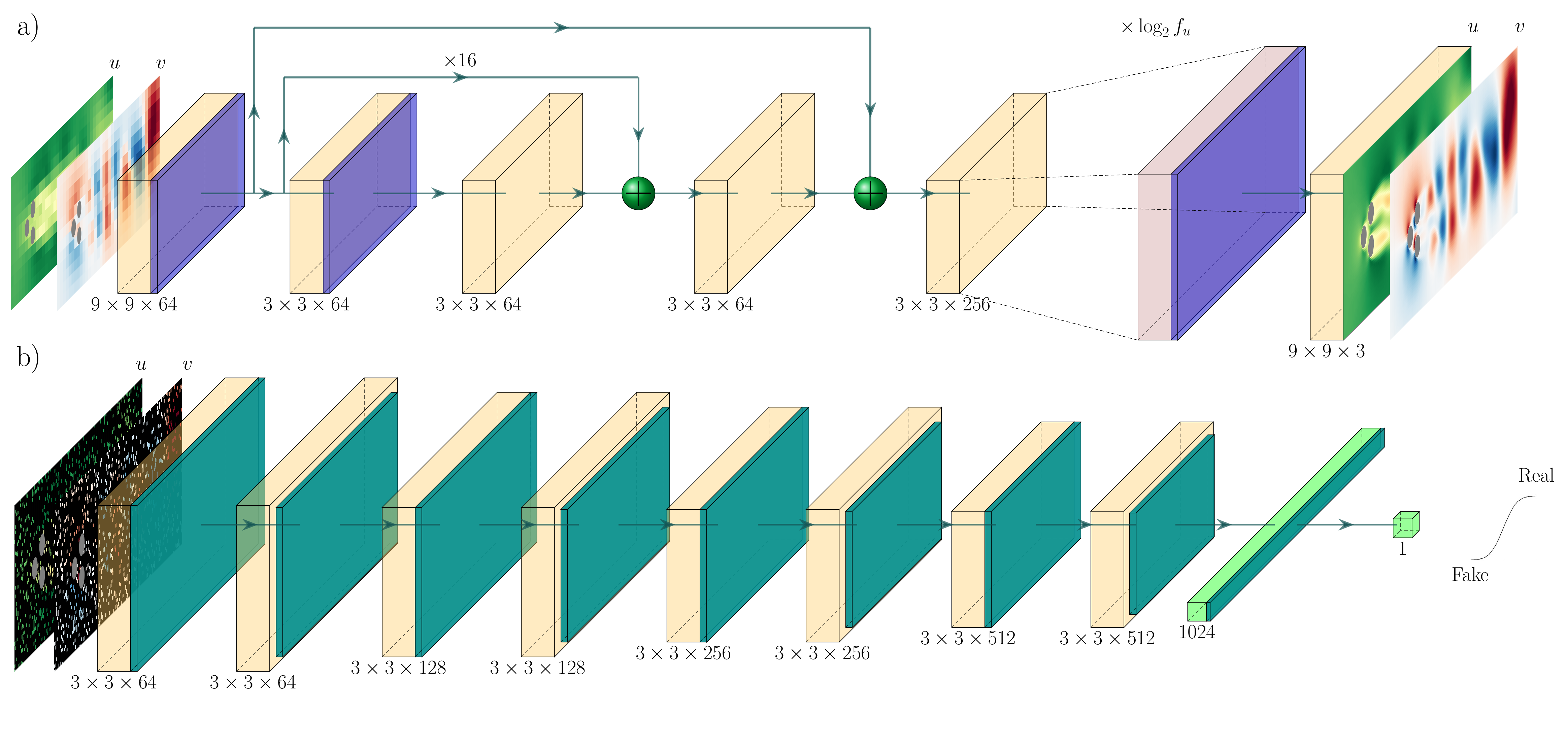}
	 \caption{Schematic illustration of RaSeedGAN architecture. a) Generator. b) Discriminator. The novel element is the use for training and testing of gappy high-resolution fields provided directly by binned scattered data.}
        \label{fig:01}
\end{figure}

Generative adversarial networks are composed of two competing networks: a generator that produces an artificial output that mimics reality, and a discriminator, which oversees distinguishing between reality and artificial outputs. For this work, the architecture proposed in \cite{ledig2017photo} has been used as a baseline. The RaSeedGAN architecture is sketched in Figure \ref{fig:01}, \textcolor{black}{and briefly outlined here. For further details on the architecture and on the training settings see the Methods section}. 

The generator is fed with full-field information at low-resolution, directly obtained from the scattered data, \textcolor{black}{and provides in output full high-resolution fields}. Several options can be explored \textcolor{black}{to generate the low-resolution fields}, including interpolation, Gaussian process regression, or Voronoi tessellation, among others. In this work, we adopt a simple binning procedure. Each bin is a compact support function whose value is assigned as the average quantity measured by the sensors within it. The bin size is defined to have approximately $10$ sensors per bin on average. This approach has several advantages, such as noise robustness, spatial resolution uniformity, and it generates an output similar to techniques used for field quantity estimation based on particle-image measurements, such as particle image velocimetry/thermometry. 

Real and generated fields are then fed into the discriminator network. 
It must be noted that the real fields are sparsely sampled, thus the full-field information provided by RaSeedGAN is masked with 0s in the empty bins of the reference \textcolor{black}{(we refer to it as ``high-resolution'')}. 

The target is built from the scattered distributions, using a binning approach similar to the proposed above. Reducing the bin size will progressively increase the probability of finding empty bins, i.e., bins without sensors. The error during training is weighted to consider only those bins in the target carrying data. 

Random sampling in space, with different distributions among realizations, is required to obtain a complete mapping of the domain of interest, i.e. RaSeedGAN requires random moving sensors covering the region of interest to be reconstructed. A single RaSeedGAN trained model can handle different sparsity levels, but the sparsity level would influence the quality of the obtained reconstructions, similarly to other methods such as cubic interpolation. In particular, in regions where there is not enough information to calculate the values corresponding to a bin location for the equivalent Low Resolution image throughout the dataset, there will be a blanked region, as in methods such as Gappy POD. Furthermore, when stacking quantities in the input of RaSeedGAN using channels in the neural network, such as can be the different velocity components $U$, $V$, their sparsity level should be similar. Better results are obtained by training a single RaSeedGAN model per quantity of interest in case of significant difference in sparsity level, although at the expense of the computational cost.

\section*{RESULTS}

\subsection*{Test cases outline}
In this section, the RaSeedGAN algorithm is applied on synthetic test cases generated from Direct Numerical Simulation (DNS) of fluid flows and on real experimental data. Two synthetic test cases with different challenges are considered: 

\begin{itemize}
    \item velocity data of the wake of a cluster of three cylinders with equal diameter $D$, referred as fluidic pinball \citep{deng2018low}. The wake of the pinball includes the interaction of the wakes of three bodies, a region of development and the final merging in a large-scale shedding wake. The domain is discretized with $25$pixels/$D$. The fields are seeded with $0.01$ vectors per pixels. The low-resolution input is built setting the binning region to $32\times32$ pixels with $50\%$ overlap, corresponding to approximately $10$ vectors per bin. Beware that these parameters aim to simulate the typical processing of PIV, in which the measured velocity closely relates to the average velocity measured within the interrogation window (here referred to as bin).
    \item a turbulent channel flow, generated from a dataset available at the Johns Hopkins Turbulence Database (\url{http://turbulence.pha.jhu.edu/}). Sub-domains are taken of size $2h\times h$ in the streamwise and wall-normal directions, where $h$ is the half-channel height. The domains are discretized with $512$ pixels$/h$. The same vector density and bin averaging size/procedure of the first test case are adopted.
\end{itemize}

In the synthetic test cases, the binned distributions for training are generated using directly the original positions on the particles (or point sensors) rather than performing particle tracking on virtual images. Noise in the measurement is introduced by adding a random perturbation on velocity vectors of realistic intensity. For the pinball test case, two Gaussian noise conditions are added to the displacement vectors to study the robustness of the method against noisy measurements, with standard deviation in pixels of $0.01$ (moderate intensity, Test Case 1a) and $0.05$ (high intensity, Test Case 1b), respectively. For the channel, a higher noise is introduced ($0.1$ pixels) to account for the higher freestream displacement if compared to the pinball case. In both cases, the error realistically represents noise perturbation of Particle Tracking Velocimetry data. With this approach, we attempt to isolate the errors due to spatial resolution limitations while maintaining a controlled random error level due to uncertainty in feature matching. 

Additionally, two complex experimental test cases are chosen:
\begin{itemize}
    \item temperature data from the NOAA sea surface temperature (\url{http:
//www.esrl.noaa.gov/psd/}), based on temperature data collected by satellites and ships. The temperature data are randomly seeded to provide artificial point sensors, while the original data are used for assessment.
    \item in-house experimental PIV data of a zero-pressure-gradient (ZPG) turbulent boundary layer (TBL). In order to have a ground truth available for comparison, the low-resolution field is estimated on a fictitiously large window of $128 \times 128$ pixels with $50\%$ overlap with a multi-step PIV interrogation \cite{scarano2001iterative}. Then an analysis using a window with $32 \times 32$ pixels is used to generate a high-resolution reference. It must be remarked here that these high-resolution fields are not used in any form in the training process and are just exploited for performance assessment. The number of vectors is then artificially reduced by a factor of 10 by randomly dropping sensors, and fed to RaSeedGAN to test the super-resolution capabilities.  
\end{itemize} 

A more detailed description of the test cases is reported in the section Methods.

The RaSeedGAN seeks to achieve upsampling levels ranging from $f_u=2-8$, with $f_u$ being the upsampling factor. The window used to bin the vector fields has a size equal to $b=D_I/f_u$, with $b, D_I$ being the bin size of the binned data and the bin size for the low-resolution data, respectively. Furthermore, the grid is refined by the same amount, thus maintaining the overlap between adjacent bins. In all cases, $D_I$ is set in order to have on average 10 sensors per bin, thus ensuring negligible probability of empty bins within the low-resolution input. It is important to underline that selecting a smaller bin $b$ increases the levels of gappyness of the distributions to be reconstructed.


It must be remarked that, in all test cases, the position of the sparse sensors on each individual snapshot is fixed during training, i.e. each individual field is sampled at a predetermined finite number of locations which does not change during training. The positions of the sensors, on the other hand, change from snapshot to snapshot. This framework is intended to replicate realistic scenarios of optical particle-based measurements, or of field measurements with moving point sensors.

\subsection{Validation on synthetic test cases}

\begin{figure*}[h!]
  \begin{center}
  \includegraphics[width=\linewidth]{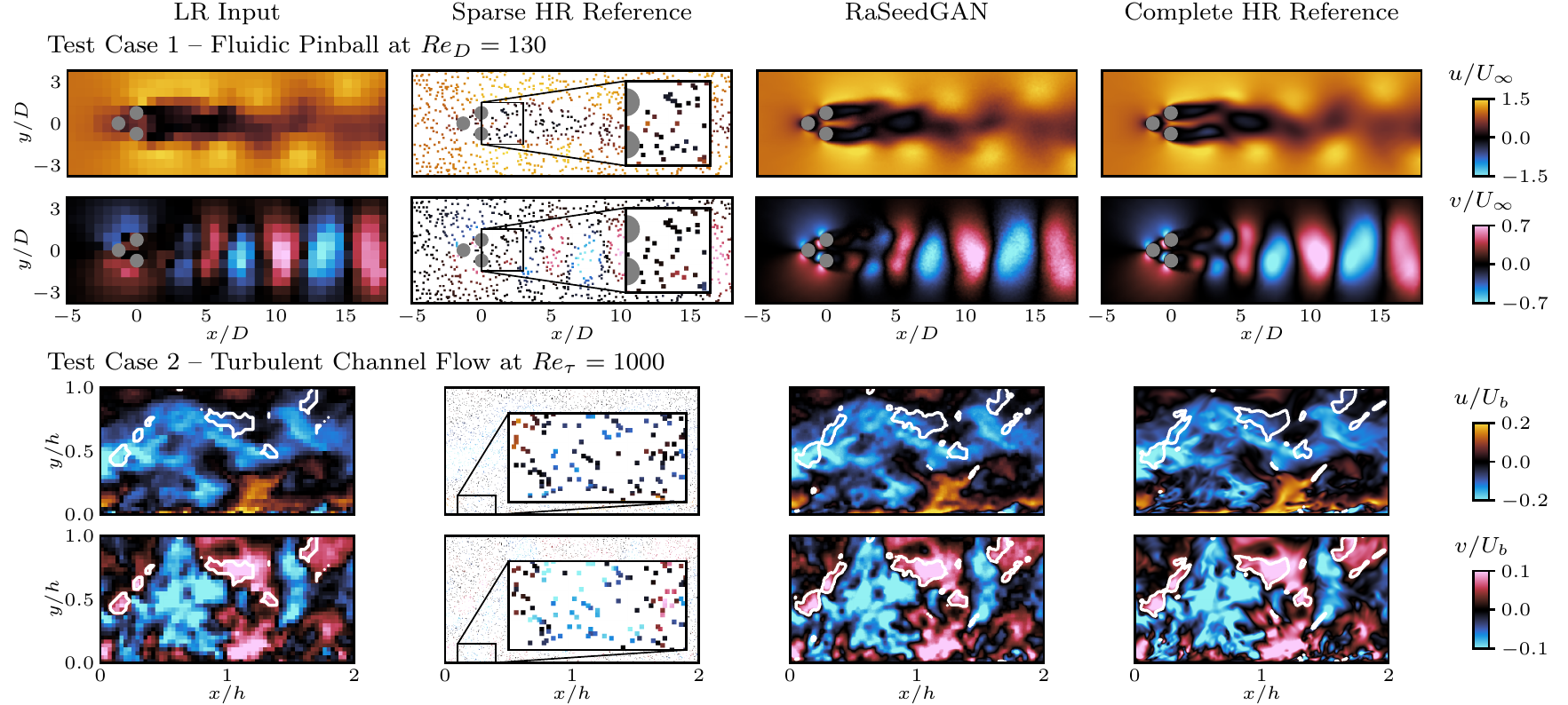}
  \end{center}
  \caption{Panel of synthetic test cases, with examples of instantaneous field reconstructions. Lower-case letters $u$ and $v$ indicate respectively the streamwise and crosswise velocity fluctuation components. The reference velocity $U_\infty$ and $U_b$ are the freestream and the bulk velocity for the pinball and the channel flow test case, respectively. $Re_D$ and $Re_\tau$ indicate the Reynolds number based on the cylinder diameter $D$, and the friction-based Reynolds number. The columns indicate, from left to right: low-resolution input field; binned distribution with $f_u=8$; reconstruction using the proposed sparsely-trained super-resolution GANs; the reference full field. From top to bottom: wake of the fluidic pinball from DNS, streamwise and crosswise velocity fluctuation components; turbulent channel flow from DNS, wall-parallel and wall-normal velocity component. White isolines in Test Case 2 indicate \textit{uvsters}. For clarity, an inset view has been included in the Sparse HR Reference to show how the sparse images are represented.}
\label{fig:02}
\end{figure*}

For all validation cases, we explore the performances of RaSeedGAN with instantaneous visualisations of reconstructed fields. Figure \ref{fig:02} shows an instantaneous realisation for the synthetic test cases considered under study, including in columns: the input low-resolution (LR) field, the binned field, referred as ``Sparse HR (high-resolution) Reference'', the reconstruction achieved with the RaSeedGAN algorithm and the reference field used to generate the particle distributions. The complete high-resolution data is included in the last column of Figure \ref{fig:02} as a quality check, although it is not used at any moment during training.
Furthermore, Table \ref{tab:table3} reports the pixel-based root mean squared error for each test case and upsampling levels compared to the error obtained by a standard cubic interpolation method \textcolor{black}{and the Gappy POD, following the implementation from Ref.~\cite{venturi2004gappy}.} The error is normalised to the standard deviation of the corresponding quantity. 

\begin{table*}[t]
\caption{\label{tab:table3} Root mean squared error for each test case, scaled with the \textcolor{black}{spatially-averaged} standard deviation of their \textcolor{black}{corresponding} fluctuating quantities. \textcolor{black}{Cases correspond to: (1a) Pinball with $1\%$ noise disturbance, (1b) Pinball with $5\%$ noise disturbance, (2) Turbulent Channel Flow with $1\%$ noise disturbance, (3) NOAA sea surface data and (4) Experimental Turbulent Boundary Layer. In bold are highlighted best-performing method for each test case.  }}
\begin{ruledtabular}
\begin{tabular}{cccccccccccc}
 Method & $f_u$  & \multicolumn{2}{c}{Test Case 1a} & \multicolumn{2}{c}{Test Case 1b} & \multicolumn{2}{c}{Test Case 2} & \multicolumn{2}{c}{Test Case 3} & \multicolumn{2}{c}{Test Case 4 \footnote{In Test Case 4 Gappy POD has been trained on a smaller dataset (5000 images) for computational cost limitations. However, only marginal improvements have been observed when increasing the dataset size.}}\\
 & & $u$ & $v$ & $u$ & $v$ & $u$ & $v$ & \multicolumn{2}{c}{$T$} & $u$ & $v$
 \\ \hline
 RaSeedGAN & \multirow{3}{*}{ 2} & 0.261 & 0.243 & \textbf{0.217} & \textbf{0.200} & \textbf{0.117} & \textbf{0.303} & \multicolumn{2}{c}{0.058} & \textbf{0.081} & \textbf{0.316}\\
  Cubic &  & \textbf{0.218} & \textbf{0.171} & 0.241 &  0.232 & 0.126 & 0.337 &  \multicolumn{2}{c}{\textbf{0.054}} & 0.176 & 0.605\\
   Gappy POD & & 0.274 & 0.243 & 0.287 & 0.277 & 0.137 & 0.373 & \multicolumn{2}{c}{0.063} & 0.172 & 0.610\\
    \hline
RaSeedGAN & \multirow{3}{*}{ 4} & 0.169 & 0.182 & \textbf{0.137} & \textbf{0.157} & \textbf{0.116} & \textbf{0.307} & \multicolumn{2}{c}{0.055} & \textbf{0.071} & \textbf{0.275}\\
  Cubic &  & 0.219 & \textbf{0.177} & 0.241 & 0.237 & 0.124 & 0.337 & \multicolumn{2}{c}{0.058} & 0.175 & 0.590\\
   Gappy POD &  & \textbf{0.163} & 0.191 & 0.180 & 0.224 & 0.163 & 0.529 & \multicolumn{2}{c}{\textbf{0.050}} & 0.171 & 0.649\\
 \hline
     
RaSeedGAN & \multirow{3}{*}{ 8} & 0.133 & \textbf{0.165} & \textbf{0.098} & \textbf{0.139} & \textbf{0.117} & \textbf{0.315} & \multicolumn{2}{c}{0.056} & - & - \\
  Cubic &  & 0.220 & 0.178 & 0.242 & 0.237 & 0.124 & 0.337 & \multicolumn{2}{c}{0.055} & - & - \\
   Gappy POD &  & \textbf{0.125} & 0.184 & 0.135 & 0.200 & 0.185 & 0.635 &  \multicolumn{2}{c}{\textbf{0.049}} & - & - \\
\end{tabular}
\end{ruledtabular}
\end{table*}

In the first row of Figure \ref{fig:02}, an instantaneous flow visualisation of the fluidic pinball case with upsampling factor $f_u=8$ is reported for both components of the velocity fields. The RaSeedGAN allows recovering correctly the developing regions around the cylinders in the near-wake and around the cylinders. For instance, the crosswise-velocity details in the front of the cylinders, which are barely seen in the low-resolution input fields, are nearly perfectly recovered. Similar performance is observed in the pinball wake, where the RaSeedGAN reconstructs the streamwise momentum defect underestimated by the low-resolution fields. The error level reported in Table \ref{tab:table3} for cases with upsampling levels of $f_u=2-8$ supports the quality of the overall reconstructed fields.

\textcolor{black}{Table  \ref{tab:table3} shows that the reconstruction errors for the Test Case 1a ($1\%$ noise level) for cases with upsampling levels of $f_u=4-8$  are significantly lower than the standard cubic interpolation, although the Gappy POD still appears to be superior due to the high energy content of the first POD modes of the pinball case. However, when the noise conditions are higher, as in Test Case 1b where a 5\% noise level is included, we observe that the accuracy of the cubic interpolation and the Gappy POD is degraded while RaSeedGAN can improve even further its performance. RaSeedGAN, like other GAN architectures, appears to be able to explore better and generate from the underlying data distribution in presence of noise \cite{radford2015unsupervised,sonderby2016amortised}. This is relevant when applying RaSeedGAN to experimental datasets where the noise level cannot be easily controlled. Additionally, it has to be remarked that the current architecture can be further improved using more data or layers/filters with respect to the baseline RaSeedGAN, at expense of a higher computational cost. As an example of this capability,  a test doubling the number of filters of the RaSeedGAN generator showed that for Test Case 1a and $f_u=2$ the pixel-based root mean squared error decreases up to $u=0.215$ and $v=0.186$.}



In the second row of Figure \ref{fig:02}, the streamwise and wall-normal velocity contours for the turbulent channel flow simulation are represented. Alongside this, contours that identify extreme Reynolds-shear-stress events - the so-called \textit{uvsters} as defined in \citep{lozano2012three} - have been plotted. The \textit{uvsters}, together with the clusters of vortices, play a crucial role in the modelling of the dynamics of wall-bounded turbulence based on coherent structures \citep{lozano2012three}. The recovery of wall-attached vertical velocity fluctuations in the near-wall region is remarkable, where even the recovery of small structures is appreciable. Regarding the outer part of the flow, the \textit{uvsters} are largely improved. Analysing the error reconstruction of Table \ref{tab:table3}, the RaSeedGAN still outperforms the standard cubic interpolation, although in this case to a lower extent due to the richer range of spatial scales involved. \textcolor{black}{
In this complex case, Table \ref{tab:table3} shows that RaSeedGAN outperforms the standard cubic interpolation and the Gappy POD for all the reported upsampling levels.}


\subsection{Application on experimental test cases}

\begin{figure*}[h!]
  \begin{center}
  \includegraphics[width=\linewidth]{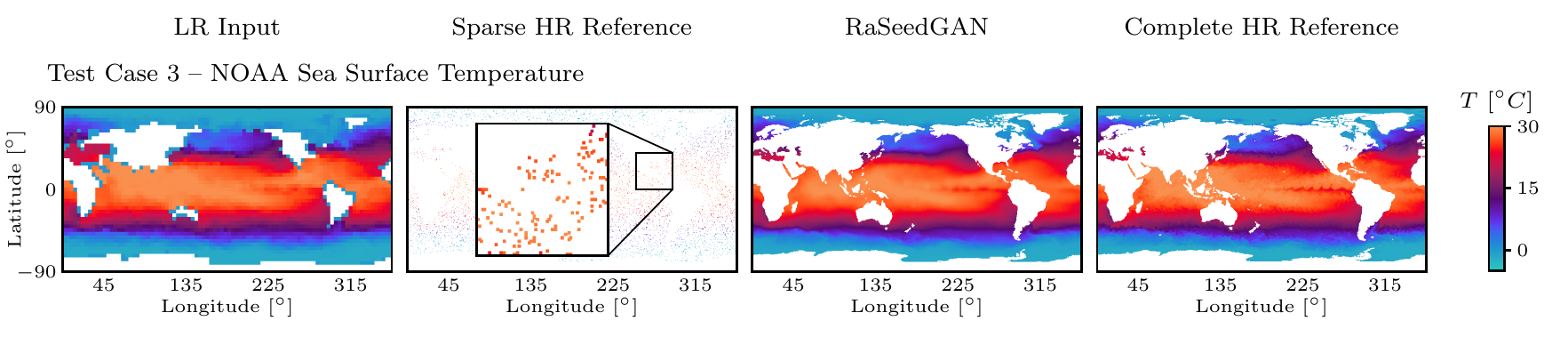}
  \end{center}
  \caption{Measurement of ocean surface temperature from NOAA sea surface data. From left to right: low-resolution input field; binned distribution with $f_u=8$; reconstruction using the proposed sparsely-trained super-resolution GANs; the reference full field. For clarity, an inset view has been included in the Sparse HR Reference to show how the sparse images are represented.}
\label{fig:testcase3}
\end{figure*}

\textcolor{black}{In this section an assessment of RaSeedGAN on the two experimental test cases is conducted.}
In Figure \ref{fig:testcase3}, an instantaneous temperature field visualisation of the NOAA sea surface temperature with upsampling factor $f_u=8$ is reported. The RaSeedGAN allows recovering the temperature fields with a higher level of detail, as is evident from the reconstructed silhouette of the continents of the Earth. \textcolor{black}{For this case the performances of RaSeedGAN seem in line with the cubic interpolation, and slightly inferior to Gappy POD. These can be ascribed to the seasonality of the data, which delivers a very compact POD eigenspectrum. However, from visual inspection and analysing the temporal spectra, it has been observed that the noise level of the RaSeedGAN is slightly lower compared with the Gappy POD and the standard cubic interpolation.}

\begin{figure*}[h!]
  \begin{center}
  \includegraphics[width=\linewidth]{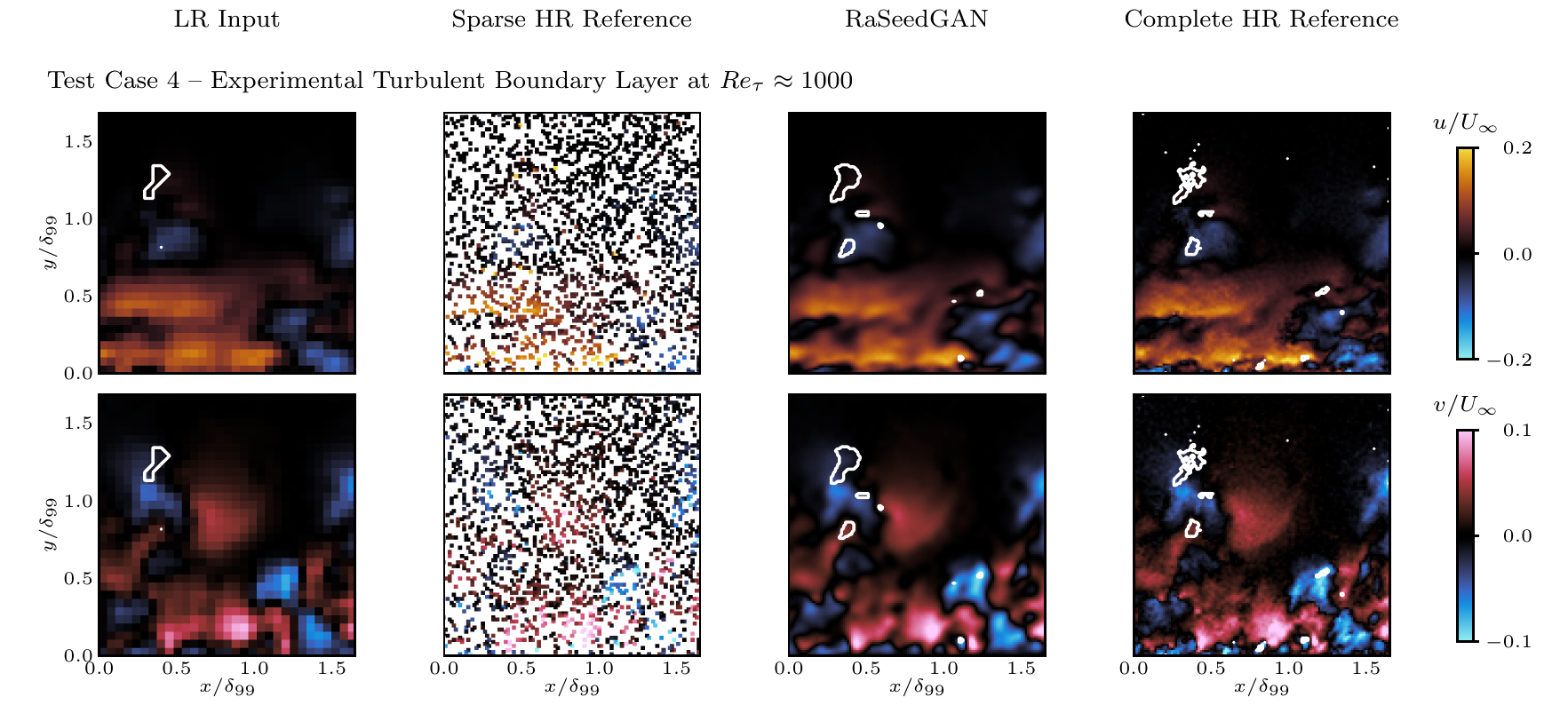}
  \end{center}
  \caption{The columns indicate, from left to right: low-resolution input field; binned distribution with $f_u=4$; reconstruction using the proposed sparsely-trained super-resolution GANs; the reference full field. The quantities $\delta_{99}$ and $U_\infty$ indicate respectively the boundary layer thickness and the freestream velocity. From top to bottom: experimental data of a turbulent boundary layer, wall-parallel and wall-normal velocity fluctuating components. Black isolines indicate \textit{uvsters}.}
\label{fig:testcase4}
\end{figure*}

In Figure \ref{fig:testcase4}, the streamwise and wall-normal velocity contours for the experimental data of ZPG TBL flow are presented. As for Test Case 2, the contours that represent the \textit{uvsters} have also been plotted. It has to be remarked that this particular case is an actual experiment in which the noisy conditions are natural and not simulated nor imposed. Again, as in the turbulent channel flow case, the recovery of wall-attached vertical velocity fluctuations in the near-wall region is observed even though the quality of the reconstruction does not match the one reported in Test Case 2. This result is not surprising since the quality of the experimental data is lower than the synthetic data. This statement is supported by the performance of the standard cubic interpolation and the Gappy POD reported in Table \ref{tab:table3}, being the difference of the RaSeedGAN and both methods more considerable than in Test Case 2.

\begin{figure*}[t]
  \begin{center}
  \includegraphics[width=\linewidth]{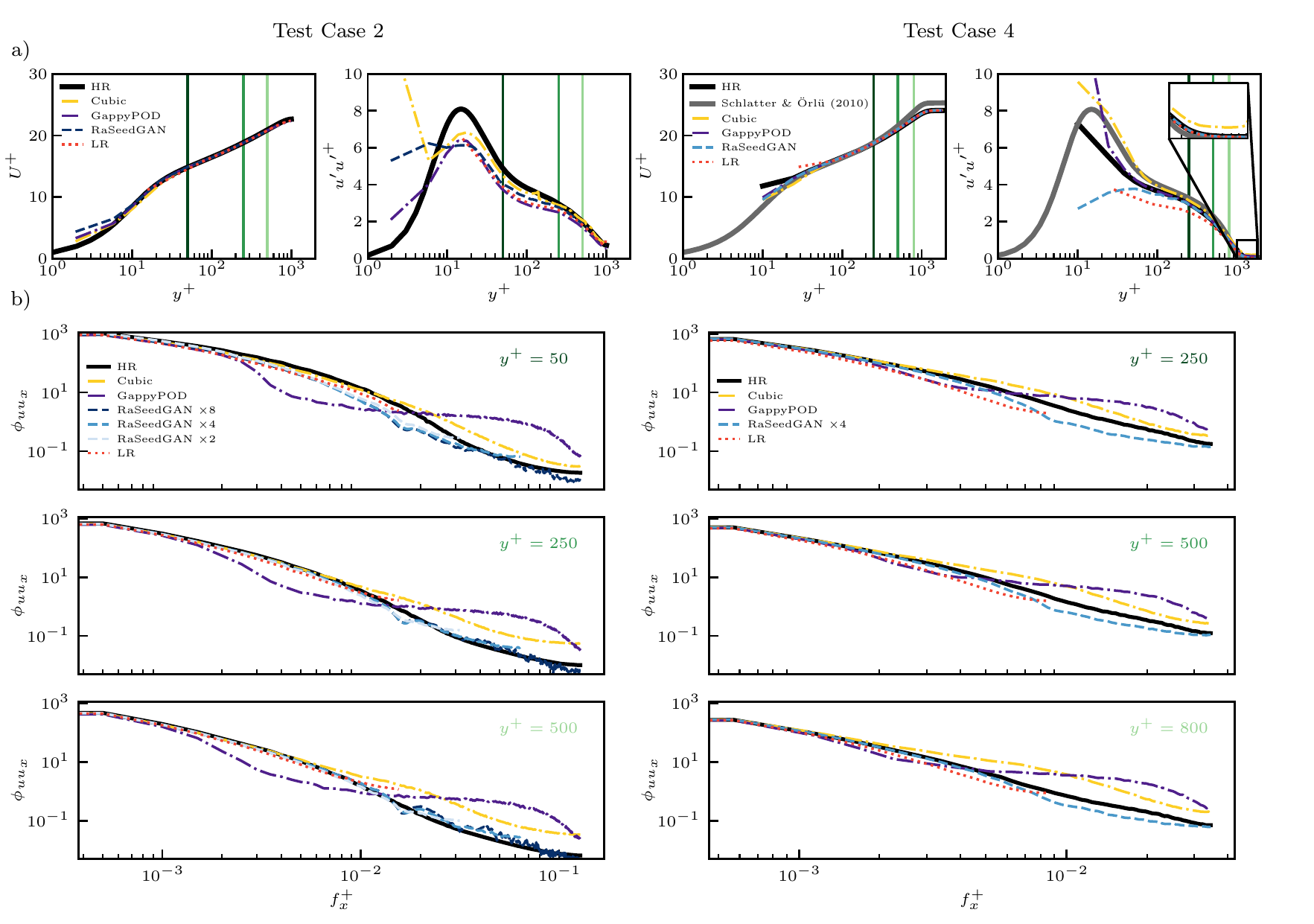}
  \end{center}
  \caption{a) Inner-scaled mean streamwise velocity and variance profiles for Test Case 2 (DNS turbulent channel flow) and Test Case 4 (experimental turbulent boundary layer). Lines represents the data used to calculate the quantity: high-resolution reference (solid black), low-resolution original measurements (dotted red), cubic interpolation (yellow dashed-dotted), Gappy POD (purple dashed-dotted) and RaSeedGAN reconstruction (dashed blue, from light to dark indicating $f_u=2,4,8$). The inset in the variance profile of Test Case 4 provides a magnified view of the region with wall distance above 1000 inner units. Vertical green lines (from dark to light for increasing distance from the wall) indicate stations where the streamwise turbulent spectra are computed. For Test Case 4, gray profiles refer to data from \citep{schlatter2010assessment}. b) Streamwise turbulent spectra at different wall distances.}
\label{fig:05}
\end{figure*}

As a further indicator of the quality of the reconstruction achieved with RaSeedGAN, the wall-normal profiles of first and second-order statistics of the streamwise velocity are reported in Figure \ref{fig:05}a. The quantities are presented in inner scaling, i.e. the mean velocity $U$ and the standard deviation of the fluctuation $u'$ are scaled with the friction velocity $u_\tau$, while the wall-normal coordinate $y$ is scaled with the viscous length scale $\ell^*$. Inner-scaled quantities are indicated with the superscript $+$. The results are obtained with ensemble averaging over the testing dataset and spatial averaging along the streamwise direction. For Test Case 2, the reference profiles are directly extracted from the simulation data in the Johns Hopkins Turbulence Database. For Test Case 4, the reference profile from the high-resolution PIV analysis is reported. For completeness, profiles from a friction-Reynolds-number matched simulation are also included \cite{schlatter2010assessment}. It is important to remark that the network is not trained to recover flow statistics but rather to reconstruct field features. \textcolor{black}{In this scenario, high-resolution statistics can be obtained with higher fidelity using directly ensemble averaging approaches on the scattered data.} \textcolor{black}{Furthermore, physics-based constraints are not included during training, i.e. RaSeedGAN in the present implementation is unaware of the underlying physics. Even though physics can be easily embedded in the training process (for instance as an additional content loss contribution based on residual from governing equations), this is not explicitly enforced in the analysis.} Consequently, this assessment can be considered a fair metric for faithful flow reconstruction.

Although RaSeedGAN seems to improve the turbulence statistics profiles of the initial low-resolution data, it is observed that the cubic interpolation seems to have better performances, especially in the inner region, i.e. below 15 wall units (which corresponds to about 8 pixels in Test Case 2, and  16 pixels in Test Case 4). This result, apparently at odd with the error statistics of Table \ref{tab:table3}, is not surprising since it is indicative that RaSeedGAN might inherit part of the intensity modulation error from the low-resolution field, while cubic interpolation works directly on the pointwise data. On the other hand, fields obtained from cubic interpolation have visibly higher noise contamination. Thus, the variance estimation might be biased towards higher intensity due to noise, partly compensating for modulation issues and providing a misleading offset towards the reference profiles. This offset effect is well known in the fluid mechanics' community \cite{atkinson2014appropriate}, and evidences of it can be observed at wall distances larger than 1000 wall-units for Test Case 4, where a plateau of the variance is observed for the cubic interpolation data to a higher value than the reference and the RaSeedGAN data (see inset of Figure \ref{fig:05}a). 
To better assess the resolution in terms of scales of the flow and discern the noise contamination, the streamwise spectra of the velocity fluctuations are computed at different distances from the wall (see Figure \ref{fig:05}b). It is observed that, except for small wall distances, RaSeedGAN can significantly increase the reach of the spectrum and follow with high fidelity the high-resolution reference. In all cases, the cubic interpolation data deliver a spectrum that peels away from the reference already at a relatively large scale. 
If considering as cut-off the frequency at which the spectrum of the low-resolution data and RaSeedGAN reach 80\% of the intensity of the reference spectrum, scales of at least 3 times smaller are recovered in all cases of Figure \ref{fig:05} except for $y^+=50$ of Test Case 2.

\section*{CONCLUSIONS}

A novel deep-learning approach based on generative adversarial networks to perform super-resolution reconstruction of sparse measurements has been proposed and assessed. The method exploits the ability of the neural networks to learn mapping functions even when some layers have randomly-ignored neurons during the training process. We propose the RaSeedGAN method based on a GAN architecture that can obtain a non-linear mapping to high-resolution reconstructed images, using as inputs sparse low-resolution images obtained by binning the randomly-distributed sensor inputs, and as targets the original incomplete sparse measurements. The proposed technique requires randomly-moving sensors from snapshot to snapshot to establish a mapping in the domain of interest. This situation is common, for instance, in particle-based optical measurements.

RaSeedGAN allows the user great flexibility for application to experimental data in configurations where high-resolution benchmarks are not available. Its main novelty is that a full field is not required beforehand and is constructed directly using gappy binned images and the original sparse measurements. Different examples that include numerical and experimental data from velocity and temperature measurements were successfully tested, demonstrating the accuracy and robustness of the proposed method even in the presence of noisy data. Even though RaSeedGAN is not explicitly trained to fulfil physical constraints, evidence of consistency with physics is found from the analysis of derived quantities, such as turbulence statistics. Furthermore, the robustness of the proposed architecture is proved in a range of cases, although the user might still find room for improvement with the tuning of the model parameters and of the loss function to minimize during training. We foresee further performance improvements from the embedding of physics laws and/or models within the training process. 

\section*{Methods}

\subsection*{RaSeedGAN implementation details}

The RaSeedGAN architecture, sketched in Figure \ref{fig:01}, is based on the one proposed in \cite{ledig2017photo}. However, the batch-normalization layers have been removed to ensure a proper fitting of the deep layers in the architecture as proposed by \cite{WangGAN}.
The generator is fed with full-field information at low-resolution, and is trained to provide in output full high-resolution fields with fixed upsampling. 

The low-resolution fields are fed to the generator, which applies a convolutional layer with filter size $9\times9$ and $64$ feature maps, followed by a parametric ReLU (PReLU) activation function. 
After these initial layers, $16$ residual blocks are applied with the layout proposed by \cite{gross2016training}, i.e., a convolutional layer followed by a PReLU activation function and a second convolutional layer, with $64$ feature maps of filter size $3\times3$ for both convolution operations. Before increasing the resolution, a skip-connection sum is performed between the residual blocks' output and the initialization layers' output. The subpixel convolution layer proposed by \cite{shi2016real} is used to increase the resolution. Finally, a convolution operation with linear activation function and filter size $9\times9$ is applied to recover the in-plane high-resolution measurements as output.

The discriminator network is then fed with real (sparse) and generated fields. The initialization of the network is carried out by a convolution operation of filter size $3\times3$ and $64$ feature maps, followed by the Leaky ReLU (LReLU) activation function. 
After this, $7$ discriminator blocks of successive increasing feature maps are applied, where the discriminator blocks are composed of convolutional layers of $3\times3$ size and LReLU activation function. 
Note that odd blocks applied a stride step larger than one to reduce the height and width of the feature maps. 
Finally, the feature-map tensor is flattened into a vector, followed by a fully-connected layer of $1024$ neurons with LReLU activation function and output fully-connected layer with one neuron and sigmoid activation function. The output of the discriminator provides the probability of whether the input is real (0s) or fake (1s).

The network architectures are implemented in the open-source framework Tensorflow \cite{abadi2016tensorflow}.
All models are randomly initialized.
Networks' weights are updated with a learning rate of $\epsilon=0.0001$.

The discriminator loss function is defined as:
\begin{equation}
    \mathcal{L}_D=-\mathbb{E}[\log D(H_R)] - \mathbb{E}[\log(1-D(F_V\otimes G(L_R)))],
\end{equation}

\noindent \noindent where $\mathbb{E}[·]$ represents the operation of taking average in the mini-batches, $H_R$ indicates high-resolution images, $L_R$ low-resolution images, $D()$ is the discriminator network that takes as input matrices with dimensionality equal to the final high-resolution grid and $F_V$ is the matrix that indicates if a bin has a sensor (1) or not (0).

The loss function of the generator network is defined as:
\begin{equation}
    \mathcal{L}_G=\sum_{i=1}^{Nx}\sum_{j=1}^{Nz}|H_{R_{i,j}}-F_V\otimes G(L_R)_{i,j}|^2+\lambda\mathcal{L}_A,
\end{equation}
where $G()$ is the generator network that takes as input matrices with dimensionality equal to the low-resolution data and $\lambda$ is a scalar parameter set to 0.001 to weigh the contribution of the adversarial loss $\mathcal{L}_A$, defined as:
\begin{equation}
    \mathcal{L}_A = - \mathbb{E}[\log(D(F_V\otimes G(L_R))].
\end{equation}

The adversarial loss is defined as the binary cross-entropy of the predicted fields, i.e., it checks whether the discriminator has labelled as `fake' the generated fields. The discriminator loss is the mean value between the binary cross-entropy of the predicted fields and the target ones. 
The mask $F_V$ is formally equivalent to a dropout layer at the last stage of the networks with random activations.

The test cases included in this work have been trained during $100$ epochs with $8$ samples per batch. The generator and discriminator states have been stored every $5$ epochs to find the optimum point before the GAN starts overfitting. 
The weights of both networks have been updated using Adam optimizer with a learning rate of $0.0001$. All the cross-entropy errors have been perturbed with a random fluctuation of standard deviation $0.2$ to increase the stability of the GAN training.

\begin{figure*}[t]
  \begin{center}
  \includegraphics[width=0.9\linewidth]{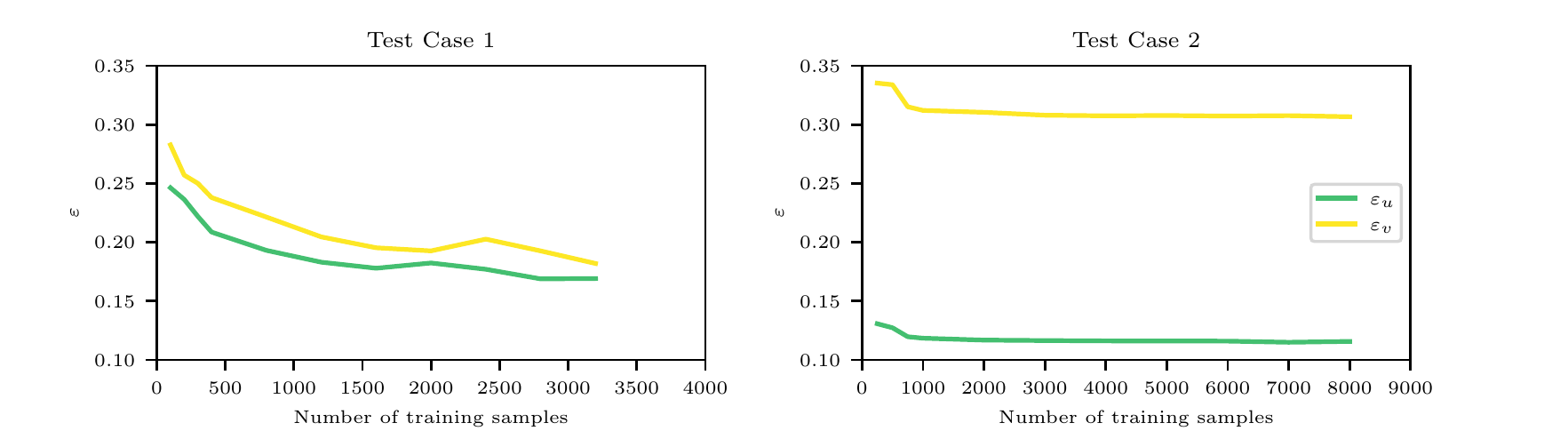}
  \end{center}
  \caption{Pixel-based root mean square error as a function of the number of training samples for Test Case 1 (left) and Test Case 2 (right) with $f_u=4$. The error is scaled with the standard deviation of the corresponding fluctuating quantities. Green and yellow lines indicate respectively the error on the streamwise and crosswise velocity components.}
\label{fig:04}
\end{figure*}

It must be remarked that we used for training a relatively large number of samples to eliminate doubts on the training convergence. Nonetheless, the performances seem to be affected only to a small extent when working on reduced-size datasets, as observed in Figure \ref{fig:04}. Test Case 1 involves a narrower range of scales and a less complex behaviour; this allows a continuous improvement of the error as the number of samples increases. On the other hand, due to the more complex behaviour of moderate-Reynolds-number channel flow, there seems to be no significant gain in increasing the number of samples beyond $1000$ for Test Case 2. It has to be remarked that the same neural network architecture has been employed for the training of every test case, and while there are complex cases such as the turbulent channel that may benefit from complex neural network architectures, it is left to the user to decide what level of complexity is required for each application.

\subsection*{Synthetic test case settings}

Test Case 1 consists of a \textit{fluidic pinball} \citep{deng2018low}, a two-dimensional wake flow around a cluster of three equidistantly-spaced cylinders with equal radius $R=D/2$, whose centres form an equilateral triangle with a side length equal to $3R$. The triangle is oriented with an upstream vertex and the downstream side orthogonal to the freestream flow, located at $x=0$ and centred on the $y-$axis. The DNS data at $Re_D=130$ (i.e. within the chaotic regime \cite{deng2018low}, \textcolor{black}{with $Re_D$ being the Reynolds number}) are used to generate synthetic PIV data. The details of the simulation settings and flow behaviour can be found in \cite{deng2018low}.

A set of $4737$ fields with scattered vectors are generated. \textcolor{black}{The simulation covers approximately $9000$ convective times (defined as $D/U_\infty$, with $U_\infty$ being the freestream velocity). This corresponds to about $1250$ shedding cycles. The snapshots have a fixed time separation of $1.9$ convective times, i.e. approximately $0.27$ shedding cycles. This allows covering correctly all phases of the shedding}. The first $4000$ fields are used for training, while the rest is used for testing. The domain spans from $-5D$ to $18.04D$ in the streamwise direction and $-3.84D$ to $3.84D$ in the spanwise direction. In order to simulate a typical PIV setup, we discretized the domain with $25$ pixels/$D$, which results in images with $576\times 192$ pixels. Approximately $1100$ vectors are generated for each sample, \textcolor{black}{corresponding to 0.01 vectors per pixel}. For this first test case, the RaSeedGAN seeks to achieve upsampling levels ranging from $f_u=2-8$, starting from an initial bin size of $D_I=32$ pixels. The level of gappyness is approximately $11\%,57\%,87\%$ for $f_u=2,4,8$, respectively.

Test Case 2, generated from a dataset available at the Johns Hopkins Turbulence Database (\url{http://turbulence.pha.jhu.edu/}), is a turbulent channel flow with a dimension of 2 half-channel-heights $h$ from wall to wall, $3\pi h$ in the spanwise direction and $8\pi h$ in the streamwise direction.
The details of the simulation settings and flow behaviour can be found in \cite{JohnsHopkinsTD}. Sub-domains of $2h\times h$ in the streamwise and wall-normal directions have been discretized with $512$ pixels$/h$ and seeded with particles with a density of $0.01$ particles per pixel. The particles have been randomly distributed in the subdomain and tracked for $10$ time steps of the simulations to generate the position in the second frame. A time separation of $1$ convective time between snapshots \textcolor{black}{(based on the bulk velocity $U_b$ and the half-channel height)} is set to \textcolor{black}{reduce statistical correlation between samples}. In order to obtain a sufficient number of snapshots, data are extracted in subdomains at different locations in the streamwise and spanwise directions. The streamwise and spanwise separation between domains was equal to $2h$ and $0.25h$, respectively. A total of $11856$ snapshots have been generated, with $10000$ used for training. 
As in the previous test case, the reference low-resolution dataset is built with a bin size of $32\times32$ pixels with $50\%$ overlap. Upsampling factors up to $8$ have been tested. The gappyness level is $12\%, 43\%, 85\%$ for $f_u=2,4,8$, respectively.

\subsection*{Experimental test case description}

Test Case 3 is the NOAA sea surface temperature (\url{http:
//www.esrl.noaa.gov/psd/}), based on temperature data collected by satellites and ships. \textcolor{black}{A total of $7305$ temperature fields has been acquired, where the training data ranges from January $1^{st}$ 2000 to June $4^{th}$ 2016 and the testing set elapses from June $5^{th}$ 2016 to December $31^{st}$ 2019.}
The original dataset has been interpolated on a grid with $720 \times 1440$ points. Point temperature measurements have been extracted at approximately $10300$ locations per sample and used to generate the low-resolution training dataset and the sparse high-resolution target. For the low-resolution training dataset, a bin size of $32\times 32$ points has been selected, resulting in approximately $10$ points on average per bin. Upsampling factors of $f_u=2,4,8$ have been explored. The gappyness level correspond to $36\%$, $69\%$ and $90\%$ for upsampling factors of $f_u=2,4,8$ respectively. 

Test Case 4 is an experimental dataset of a turbulent boundary layer flow in a wind tunnel with Particle Image Velocimetry. The dataset comprises $39000$ images, with a spatial resolution of $48.4$ pixels$/mm$. The real density of identified vectors is $0.006$ vector per pixel, which results in approximately $25000$ vectors in an area of $2048 \times 2048$ pixels. The flow fields have a domain size $1.68\times 1.68$ $\delta_{99}$ in the streamwise and wall-normal directions, respectively, where $\delta_{99}$ is the boundary layer thickness. The input low-resolution field are estimated on a fictitiously large window of $128 \times 128$ pixels with $50\%$ overlap, while the reference high-resolution fields are obtained using a window with $32 \times 32$ pixels. The number of vectors is then artificially reduced to only $2560$ vectors per sample by randomly dropping sensors. This corresponds to $10$ vectors in the original window of $128\times 128$ pixels, thus resulting in a mean vector spacing of $40$ pixels. The sparse binned distributions for GANs training are obtained by binning with upsampling factors $f_u=2,4$, resulting in a gappyness level of $15\%$ and $61\%$, respectively. Of the $39000$ fields available, $30000$ are used for training, while the rest are used for testing.

\section*{Data availability}
All datasets used in this work are openly available in Zenodo, accessible at 10.5281/zenodo.7191210

\section*{Code Availability}
All codes developed in this work are openly available in GitHub, accessible through the link:\

https://github.com/eaplab/RaSeedGAN

\section*{Acknowledgements}
This project has received funding from the European Research Council (ERC) under the European Union’s Horizon 2020 research and innovation programme (grant agreement No 949085). NOAA High Resolution SST data provided by the NOAA/OAR/ESRL PSL, Boulder, Colorado, USA.

\section*{Author contributions}
AG: Methodology, Software, Validation, Investigation, Data Curation, Writing - Original Draft, Writing - Review \& Editing, Visualization. CSV: Conceptualization, Methodology, Writing - Original Draft, Writing - Review \& Editing. SD: Methodology, Software, Resources, Data Curation, Supervision, Writing - Original Draft, Writing - Review \& Editing, Funding acquisition

\section*{Competing interests}
The authors declare no competing interests.

\section*{References}

\end{document}